\documentclass[12pt,onecolumn,showpacs,amsmath,amssymb,aps,nofootinbib,floatfix]{revtex4-2}
\usepackage{graphicx} 
\usepackage{amsmath}
\usepackage{indentfirst} 
\usepackage{appendix}
\usepackage[top=3.0cm,left=3.0cm,right=2.0cm,bottom=2.0cm]{geometry}
\usepackage[utf8]{inputenc}
\newcommand{\be}{\begin{eqnarray}}
\newcommand{\wardid}{\eqn{2point1}-(\ref{2point3})}

\newcommand{\lnz}{\ln \mathcal{Z}}

\newcommand{\ee}{\end{eqnarray}}

\newcommand{\ave}[1]{\left\langle #1 \right\rangle}
 \newcommand{\eqn}[1]{Eq.\,(\ref{#1})}
\newcommand{\eqcomma}{\phantom{AA},\phantom{AA}}

\usepackage{amssymb}
\usepackage[T1]{fontenc}
\usepackage{titlesec}
\usepackage{xcolor}
\titleformat{\chapter}
  {\normalfont\fontsize{14}{14}\bfseries}{\thechapter}{1em}{}
\titleformat{\section}
  {\normalfont\fontsize{14}{14}\bfseries}{\thesection}{1em}{}
\titleformat{\subsection}
  {\normalfont\fontsize{14}{14}\bfseries}{\thesubsection}{1em}{}
\titleformat{\subsubsection}{\normalfont\fontsize{14}{14}\bfseries}{\thesubsubsection}{1em}{}

\begin{document}
\title{Gaussian pseudogauge invariant hydrodynamics with spin}
\author{David Montenegro, Mariana Julia Pereira Dos Dores Savioli, Giorgio Torrieri}
\affiliation{Universidade Estadual de Campinas - Instituto de Fisica Gleb Wataghin\\
Rua Sérgio Buarque de Holanda, 777\\
 CEP 13083-859 - Campinas SP\\
}
\begin{abstract}
Extending the Gaussian covariant hydrodynamics approach \cite{torr1} using torsion as an auxiliary field we formulate a fluctuating hydrodynamics with spin which is  covariant with respect to pseudo-gauge transformations as well as generally covariant with respect to foliations.  
This is done via the second order gravitational Ward identities, derived here in the torsionful case.
This ensures that, while angular momentum observables depend covariantly on the pseudo-gauge, the dynamics is pseudo-gauge independent, thus clarifying the role of the pseudo-gauge in hydrodynamics with spin
\end{abstract}
\maketitle
The problem of defining spin hydrodynamics \cite{spin1,push,dirk,spin2,rybreview,spinnonrel,becspin,tors1,tors2,tors3,spin3,spin4,speranza2,spinpot1,spinpot2,spinpot3,zubbec}, motivated by the experimental observation of polarization induced by vorticity \cite{spin1,push}, has proven to be challenging at a conceptual and not just at a calculational level. The combination of the concepts of spin and local equilibrium forces us to rethink the textbook assumptions under which hydrodynamics and transport theory is formulated, from isotropy \cite{spinnonrel} to the proper definition of vorticity to the separation between non-dissipative and dissipative dynamics \cite{spin4}.  
 
One important issue is that of defining and physically clarifying the role of the pseudo-gauge \cite{pseudopush,pseudogaugebuz,jain,speranza,flork,pseudogauge2} in local thermal equilibrium.  It has been known for a long time \cite{weinberg,misner,hehl,noether,jeonspin,brauner,freese} that when spin is considered, the energy momentum tensor $T_{\mu \nu}$ becomes ambiguous, under transformations of the type
\begin{equation}
\label{pseudodef}
T^{\mu \nu} \rightarrow T^{\mu \nu} +\frac{1}{2} \partial_\alpha \left( \phi^{\alpha \mu \nu}  + \phi^{\nu \mu \alpha} + \phi^{\mu \nu \alpha} \right)\eqcomma S^{\mu \nu \lambda} \rightarrow S^{\mu \nu \lambda} -\phi^{\mu \nu \lambda}  
\end{equation}
where $\phi^{\alpha, \nu \nu}$ is tensor-valued field in spacetime antisymmetric in $\nu,\mu$.  

It is obvious from tensiorial symmetry that if $T_{\mu \nu}$ is a conserved current, i.e. $\partial_\mu T^{\mu \nu}=0$, then so are the redefinitions of \eqn{pseudodef}.
Within field theory this redefinition possibility is in fact useful since the canonical energy-momentum tensor for fields transforming non-trivially under space-time transformations
\begin{equation}
T_{\mu \nu}=  \sum_{ab}\frac{\partial L}{\partial (\partial^\mu \psi_a )}\partial_\nu \psi_b - g_{\mu \nu}L 
\end{equation}
(Where ``a`` are spinorial indices if $\psi_a$ is a spinor, 4-vector components if $\psi$ is a vector and so on) are generally a-symmetric, and hence can not serve as sources for a metric-defined gravitational field.
Arbitrary redefinitions of the kind of \eqn{pseudodef} transform between equally conserved currents, only one of which (the so-called Belinfante-Rosenfeld form) is fully symmetric.  Since in hydrodynamics with spin spin is a local intensive quantity, the dynamics is generally pseudo-gauge dependent.  In particular,since the equation of state depends on spin density it generally requires asymmetric energy-momentum tensors.

   Mathematically, the ambiguity of \eqn{pseudodef} might reflect the rather trivial statement \cite{jeonspin} that a gradient of an antisymmetric tensor added to a conserved symmetric tensor is still conserved, but the physical interpretation of this is not so trivial.  As \cite{jeonspin} notes, this mathematical ambiguity is connected with the issue of localizing the physical degrees of freedom.
   Physically, pseudogauge transformations can be thought of as redistributing angular momentum between spatial angular momentum (non-extensive in space) and spin (localized in particle excitations), and might indicate that defining locally conserved currents is only unambiguously possible if additional constraints are met \cite{jeonspin};  However, this begs the question of weather dynamics of locally equilibrated fluids with spin should be pseudogauge invariant or not.   This question is intimately related to the fact that spin hydrodynamics inherently deals with the backreaction of microscopic states on the macrostate \cite{rybreview,spin2}.

   Looking at attempts to write down spin hydrodynamics in the literature, it is found that a deterministic theory determined by conservation laws and constituent relations \cite{spin1,spin2,spin3,spin4,jain} either the dynamics itself is dependent on the pseudo-gauge transformation or, at most \cite{tors1,tors2,tors3}, constitutive relations are pseudo-gauge covariant in a way that makes the dynamics pseudo-gauge invariant.
   For theories {\em not} determined by conservation laws but by the lagrangian, such as \cite{spin2,spin4,rybreview}, dynamics is pseudo-gauge invariant if one does not include fluctuations, since the Pseudogauge acts as a boundary term.  However, it is found that in most hydrodynamic theories with spin.  Even in this case, however, entropy current (which does not appear explicitly in the dynamics) is pseudo-gauge dependent, and curcially it would make fluctuations pseudo-gauge dependent.    Some researchers accept this conclusion \cite{flork,jain,pseudogaugebuz,speranza,speranza2}, arguing that pseudo-gauge is somehow physical when quantum mechanics and fluctuations are concerned, or that it is a field redefinition equivalent to frame choice in different formulations of hydrodynamics, to be fine-tuned if the effective theory is to be constructed including pseudogauge invariance.
   But, beyond the foundational issue of how a basic concept such as local equilibrium depend on a seemingly unphysical ''trick'' such as the pseudogauge, this approach is inherently suspect when spin-1 particles are considered,as in that case pseudo-gauge and gauge transformations intertwine \cite{ghosts}.

An approach to hydrodynamics in terms of a partition function and non-perturbative fluctuations constrained by symmetries, rather than as a deterministic conservation law evolution equation\cite{crooks,torr1,torr2,torr3,diffusion,grossi}, has provided a new way to think about the problem. Pseudo-gauge transformations could be ''redundancies'', analogous to ghosts in quantum field theory, which disappear provided the fundamental symmetries of the theory, namely general covariance under spacetime refoliations \cite{torr2,torr3}, are implemented.
As shown in \cite{torr3}, such a dynamics can be given by the so-called ''Gravitational Ward identity'' provided the partition function is assumed to be simple enough to be modeled by a Gaussian.

 Neglecting chemical potential (for simplicity, but it can be straightforwardly included without changing anybresults here \cite{diffusion}) the exact partition function with spin can be written down in terms of a space-time causal foliation $d\Sigma_\mu$ as \cite{zubbec}
\begin{equation}
\label{zubarevdef}
\mathcal{Z}=\mathrm{Tr}_{\psi} \exp \left[-\int_\Sigma  d\Sigma_\mu \left( \beta_\nu \left[ \hat{T}^{\mu \nu} -\frac{1}{2} \Omega^\mu_{ \beta  \alpha}\hat{S}^{\beta \nu \alpha}\right) \right]\right]
\end{equation}
where $\beta_\mu$ is the Lagrange multiplier for the momentum while $\Omega_{\mu \nu \alpha}=x_\mu T_{\nu \alpha}-x_\nu T_{\mu \alpha}$ represents the vorticity.  The tracing is done over the microscopic degrees of freedom, denoted as $\psi$.   Note that this implies that vorticity in a thermal background identified with the curl of enthalpy$\times$flow, the only one for which Kelvin`s theorem applies and the the only one associated with a conserved quantity.  

This is an extremely complicated object, a functional integral whose integrand depends on tensor and 3-form fields.  In the usual deterministic approach \cite{spin1}, one can expand derivatives of \eqn{zubarevdef} around gradients to get constitutive relations.

Following \cite{torr1}, however,
we shall expand the partition function around it's equilibrium stationary value we get, to leading order
\begin{align}
\label{lnzexpansion}
& \lnz \simeq  \left. \lnz \right|_0 - \textcolor{black}{\int_{\Sigma,\Sigma'}}
 d\Sigma_\alpha  d\Sigma'_\tau \Big[ \left. \frac{\partial^{2} \lnz}{\partial \beta_\mu \partial \beta_\nu} \right|_0  \  \left( T^{\mu \alpha}(\Sigma)-\ave{T^{\mu \alpha}(\Sigma')} \right)\left( T^{\nu \tau}(\Sigma) - \ave{T^{\nu \tau}(\Sigma')} \right) \nonumber \\ & + \left. \frac{\partial^{2} \lnz}{ \partial_\mu \beta_\nu  \partial_\beta \Omega_{\gamma \rho} } \right|_0  \left( T^{\mu \alpha}(\Sigma)-\ave{T^{\mu \alpha}(\Sigma')} \right) \left( S_{\beta \gamma \rho} (\Sigma) -\ave{S_{\beta \gamma \rho} (\Sigma') } \right) \Big] + \left. \frac{\partial^{2} \lnz}{ \partial_\mu \Omega_{\nu \alpha}  \partial_\beta \Omega_{\gamma \rho} } \right|_0  \times \nonumber \\ 
 &  \left(  S_{\mu \nu \alpha} (\Sigma) -\ave{S_{\mu \nu \alpha} (\Sigma') }\right) \left( S_{\beta \gamma \rho} (\Sigma) -\ave{S_{\beta \gamma \rho} (\Sigma') } \right) \Big].
\end{align}
which implies that the probability of a given configuration obeys
the most general Gaussian ansatz, 
\begin{align} \label{partfull}
 \mathcal{Z} \left(T_{\alpha\beta},S_{\mu \nu \alpha}\right)\simeq \prod_{\Sigma \Sigma'} & \exp [ -C^{\mu \nu \alpha \beta}(\Sigma,\Sigma') \left(  T_{\mu \nu}-\ave{T_{\mu \nu}(\Sigma)}\right) \left( T_{\alpha\beta}-\ave{T_{\alpha \beta}(\Sigma')}\right)  \nonumber \\ 
&  \qquad + Q_{\mu \nu \alpha \beta \gamma} (\Sigma,\Sigma')\left(  S^{\mu \nu \alpha}-\ave{S^{\mu \nu \alpha}(\Sigma}\right)  \left(  T^{\beta \gamma}-\ave{T^{\beta \gamma}(\Sigma')}\right)  \nonumber \\ & \qquad
- W^{\mu \nu \alpha \beta \gamma \rho} (\Sigma,\Sigma') \left(  S_{\mu \nu \alpha}-\ave{S_{\mu \nu \alpha}(\Sigma)}\right) \left( S_{\beta \gamma \rho}-\ave{S_{\beta \gamma \rho}(\Sigma')}\right)
\end{align}
For zero spin ($Q_{\mu \nu \alpha \beta \gamma}= W^{\mu \nu \alpha \beta \gamma \rho}=0$) these equations were examined in \cite{torr1}.  
The partition function \eqn{partfull} can be integrated as
\begin{align}
\lnz=\int_{\Sigma,\Sigma'} & d \lnz(\Sigma,\Sigma'), \quad  d\lnz=\left( \sqrt{
C^\prime_{\mu \nu \alpha \beta}\times\text{diag}(\lambda_{1,2,3,4}(\Sigma))\delta^{\mu \nu}\times\text{diag}(\mu_{1,2,3,4}(\Sigma'))  \delta^{\alpha \beta} }\right)^{-1} \nonumber \\
& \qquad \qquad \qquad \quad \times \left( \sqrt{ Q^\prime_{\mu \nu \alpha \beta \gamma} \times \text{diag}(\alpha_{1,2,3,4,5}(\Sigma)) \delta^{\mu\nu} \times \text{diag}(\beta_{1,2,3,4,5}(\Sigma')) \epsilon^{\alpha \beta \gamma}}\right)^{-1}  \nonumber \\ 
& \qquad \qquad \times \left( \sqrt{W^\prime_{\mu \nu \alpha \beta \gamma \rho}\times\text{diag}(\xi_{1,2,3,4,5,6}(\Sigma))\epsilon^{\mu\nu \alpha}\times \text{diag}( \chi_{1,2,3,4,5,6}(\Sigma'))\times \epsilon^{\beta \gamma \rho} }\right)^{-1} 
\end{align}
where $\mathrm{diag(...)}$s are defined by the generalization of Eigenvalues for higher rank tensors (constructed as in \cite{torr1}) and
where $A^\prime$ refers to $A$ in the frame where the average of the ensemble is diagonal.    This is ithe only way that a ''frame field'' can be defined in a generally covariant theory, and it physically reflects the difference between ''the cell'' and ''the bath'' with which the cell exchanges conserved quantities (note that away from the ideal limit the fluctuation-dissipation limit requires fluctuations, and hence baths \cite{torr1,torr2}).
While this ansatz might seem rough, and indeed as discussed in \cite{torr1} the integral measure generally leads to ``logarithmic`` non-Gaussian effects, we believe it is adequate for nearly locally thermalized fluids away from the critical point.  It is well-known that this ansatz works for strongly-coupled systems \cite{kovner} and also Gaussian distributions are the only known stable renormalization group fixed point \cite{gauss2} (a result related to the well-known Donsker`s theorem).

Then, if both the propagator of two energy-momentum tensors and the average of the energy-momentum tensor are determined by this partition function 
\begin{equation}
\label{tgdef}
  \ave{T^{\mu \nu}(x,t)}= \frac{\delta}{\left. \delta g_{\mu \nu}\right|_{x,t}} \lnz \eqcomma  C_{\mu \nu \alpha \beta}(x,x')= \frac{\delta^2}{\delta \left. g_{\mu \nu} \right|_{x,t} \delta \left. g_{\mu \nu} \right|_{x',t'} } \lnz,
\end{equation}
then, provided the partition function $\lnz$ can be approximated by the Gaussian, the Gravitational Ward identity \cite{boulware,heinz} 
together with a linear response relation \cite{forster} are enough to evolve the partition function, since they are two equations with two uknowns, $\ave{T_{\mu \nu}(\Sigma)}$ and $G_{\mu \nu \alpha \beta}(\Sigma,\Sigma')$.  
The Ward identity ensures that the evolution is generally covariant under changes of metrics, defined by refoliations $g_{\mu \nu}=\partial\Sigma_\mu/\partial \Sigma^\nu$.
One can see this as an application of Noether`s first and second theorem \cite{noether}.

Physically, one can think of this equation as evolving {\em an ensemble} defined by an average and a correlator (which in the Gaussian approximation are sufficient to specify the ensemble).
 This is reasonable since, as argued previously \cite{torr2,rybreview} spin degrees of freedom appear in hydrodynamics at the level of the fluctuation scale (as they are related to the distribution of microstates) rather than at the dissipation scale.
 
 Mathematically, this approach allows us to write the dynamics
in a way that is invariant under causal spacetime foliations:  The order of spacelike events is generally not invariant under different choices of the global foliation $\Sigma_\mu(x,t)$.  Hence, what under one foliation looks like an equal time correlator describing equilibrium fluctuations, under another it could be a linear response driven dissipative evolution. 

If there is a way to extend this approach to spin hydrodynamics, general covariance would also imply pseudogauge invariance. As shown in \cite{brauner}  physically pseudogauge transformations can also be thought as field redefinitions coupled to non-inertial transformations, an example of the application of Noethers second theorem \cite{freese} which can be implemented at the level of the partition function \cite{grossi}.   More concretely, momentum conservation in quantum field theory is implemented by \cite{misner}
\begin{equation}
  \label{eq:angmomcons}
\frac{d}{d\Sigma_0} J^{\mu \nu}=0 \eqcomma J^{\alpha \beta}=\oint  \left(\Omega^{\alpha \beta \gamma} + \ave{S^{\alpha \beta \gamma}}  \right) d^{3} \Sigma_\gamma
\end{equation}
where $\Omega^{\alpha \beta \gamma}$ is the orbital angular momentum density (vortical in a fluid, calculated with a symmetric tensor) and $S^{\alpha \beta \gamma}$ is the spin.   Looking at \eqn{pseudodef} it is clear that varying $\phi_{\mu \nu \alpha}$ redistributes angular momentum density between these two terms, which is obviously in contradiction with local entropy maximization.  
Here, the intuition of \cite{torr1} applied to local changes in the foliation $\Sigma_\mu$ turns  useful: if the dynamics is defined at the ensemble level in terms of correlators, changes in the pseudo-gauge or of $d^{3}\Sigma_\mu$ will change in general, $\Omega^{\alpha \beta \gamma},\ave{S^{\alpha \beta \gamma}}$ of {\em every element of the ensemble} but not the dynamics {\em of the ensemble as a whole} because such changes are compensated by changes in the correlator (in \cite{torr1} this meant a change in $\beta_\mu$ and the covariantized heat capacity).

The problem, of course, is that $\ave{S^{\alpha \beta \gamma}}$ is not obtainable from the partition function unless torsion  \cite{shapiro,fabbri,carroll,probe,impediment,koivisto,pereira} is considered.
In fact, spin hydrodynamics has been derived in the past \cite{tors1,tors2,tors3} using torsion.  In such an approach torsion is an auxiliary field, used to obtain the mean spin density from the partition function.  The dynamics in \cite{tors1,tors2} is pseudo-gauge invariant, but the resulting entropy current is not, which means that extending this approach to fluctuations would generate a pseudogauge dependence \cite{flork}.  Note that this does {\em not} mean that torsion is ''physically'' there, something both conjectured in some extensions of gravity \cite{shapiro,fabbri,carroll,probe,impediment} and used to rewrite general relativity in a more ''gauge-like'' form \cite{koivisto,pereira};  The fluid in a heavy ion collision \cite{spin1} lives in a space free of either Riemannian or torsional curvature.  However, following \cite{torr1} the coordinate system appropriate for hydrodynamics is not necessarily flat, and in a situation of spin density \cite{spinpot1,spinpot2} torsional degrees of freedom might be the best way to describe the situation.

To achieve this aim, one must first understand what the gravitational Ward identities (as opposed to the field theory ones calculated in \cite{tors3}), which determine the generally covariant dynamics, must look like with torsion, in other words to understand how the Ward identity and \eqn{tgdef} look once the partition function contains the sort of torsion terms in \cite{tors1,tors2}.  This is the purpose of our project.

To generate a spin density, which transforms as $\phi$ in \eqn{pseudodef}  one would need to relax the symmetry condition $\Gamma^\mu_{\rho \sigma}=\Gamma^\mu_{\sigma \rho}$ and differentiate the partition function $\lnz$ with Vierbien \cite{tors1} with the metricity condition relaxed (see eq. 3.35 of \cite{tors1})
\begin{equation}
\label{parttorsion}
\ave{T_{\mu a}}=\left.\frac{1}{\left|e\right|}\frac{\delta}{\delta e^{a \mu}}  \lnz\right|_{K_\mu^{ab}=0}\eqcomma \ave{S_{\mu ab}}=\frac{1}{\left|e\right|}\left. \frac{\delta }{\delta \omega^{ab \mu}}\lnz\right|_{K_\mu^{ab}=0}
\end{equation}
where the new integration variables reflect the torsional degrees of freedom (latin letters are comoving indices, greek are spacetime ones)
\begin{description}
\item[The Vierbein] $e_\mu^{a}$ relates the comoving Minkowski metric to the arbitrary metric $\eta_{ab}$, $g_{\mu \nu}=e_a^\mu e_b^\nu \eta^{ab}$
\item[The spin connection] $\omega_\mu^{ab}$ defines the covariant derivative of a spinful quantity
\begin{equation}
\label{torscov}
\nabla_\mu Q^{a}_\nu=\partial_\mu Q^{a}_\nu+\omega_{\mu c}^{a}Q^{c}_\nu-\Gamma^{\alpha}_{\mu \nu}Q_\alpha^{a}
\end{equation}
\item[The contorsion tensor]$K_\mu^{ab}$ is the antisymmetric term in the connection, (the torsion tensor $\mathcal{T}^\mu_{\alpha \beta} =\Gamma^{\mu}_{\alpha \beta}-\Gamma^{\mu}_{\beta \alpha}$ ) projected into the comoving coordinates. It fulfills a role similar to curvature in Riemannian manifolds 
\end{description}
Without torsion (denoted by $\breve{A}$) all of these objects are determined by the metric
\begin{equation}
 \breve{\omega}^{ab}_\mu=e_\nu^{a}\left( \partial_\mu e^{\nu b}+\breve{\Gamma}_{\sigma \mu}^\nu e^{\sigma b}\right) \eqcomma K_{\mu}^{ab}=\omega^{ab}_\mu-\breve{\omega}^{ab}_\mu \rightarrow 0
 \label{spin and contorsion}
\end{equation}

Where the spin connection comes from the tetrad postulate:
\begin{equation}
    D_{\mu}e^a_{\nu}=\partial_{\mu}e^a_{\nu}- \Gamma^{\rho}_{\mu\nu}e^a_{\rho}+\omega^a_{\mu b}e^b_{\nu}=0,
\end{equation}

 We note here that the pseudogauge as defined in \eqn{pseudodef} is related to the spin density as defined in \eqn{parttorsion} by $\phi_{\mu}^{ \alpha \beta}=S_\mu^{ab} e_b^\alpha e_a^\beta$.
The equivalent relation to the conservation law \eqn{covcons} for torsional matter is the top Eq. 2.22 of \cite{tors3} (only the first term is relevant for flat space in equilibrium).

Now we remember that, for a quantity $J$ generated from the partition function via $A$ the correlator $\ave{J(x)J(x')}$ is obtainable from the average via 
\begin{equation}  
\label{gencurrent}
\ave{J(x)J(x')} = \frac{\delta^{2} \lnz}{\delta A(x) \delta A(x')}= \frac{\delta }{\delta A(x')} \ave{J(x)}
\end{equation}
For $J\equiv T_{\mu \nu}$ and $A\equiv g_{\mu \nu}$ applying \eqn{gencurrent} to both sides of the covariant conservation of energy-momentum
\begin{equation} 
\label{covcons}
\frac{\delta}{\delta g_{\mu \nu}} \left[ \partial^\mu \ave{T_{\mu \nu}}+\Gamma_{\nu}^{\rho \sigma} \ave{T_{\rho \sigma}} \right] =0,  
\end{equation}
gives the Ward identity (Eq. 52 of \cite{heinz} ).  Note that this happens because the Christoffel symbol $\Gamma_{\mu \alpha \beta}$ depends on the metric and hence enters the derivative (no equivalent of contact terms exists in {\em vector} conserved currents, i.e. locally conserved scalar charges \cite{diffusion}), which will also be true for a gravity with torsion.

For torsion, we have two pairs of such relations: $J \equiv T_{\mu}^{a},S_{\mu}^{ab}$ and $A\equiv e_{\mu}^{a},\omega_\nu^{ab}$ respectively.    

Differentiating both sides of Eq. 2.22 of \cite{tors3} (the equivalent equation is Eq 1 of \cite{tors2}.  In both cases we only use terms relevant to spin hydrodynamics in flat space) we get (with the notation $A_{[\alpha \beta]}=\frac{1}{2}\left(A_{\alpha \beta}-A_{\beta\alpha}\right)$
\begin{equation}
\label{alldiffs}
\frac{\delta}{\delta \left[ e_\mu^{a},\omega_\nu^{ab} \right]} \left[
\begin{array}{c}
\breve{\nabla}_\mu \ave{T^{\mu \nu}}+\ave{T_{\rho \sigma}} K^{\nu ab} e^\rho_a e^\sigma_b \\
\breve{\nabla}_\lambda \ave{S_{\mu \nu}^\lambda}-2 \ave{T_{[\mu \nu]}}+2\ave{S^\lambda_{\rho [\mu} }e_{\nu ]}^{a} e^{\rho}_b K_{\lambda a}^b
\end{array}
\right]=0
\end{equation}
where in the limit of vanishing torsion  \eqn{spin and contorsion} can be used to eliminate one of the integrals, as in that case $\partial_\mu e_\nu^a=-e_\mu^b \omega^{a}_{\nu b}$.

\eqn{alldiffs} together with the two-point functions 
\begin{align}
G^{\mu a,\rho b}(\Sigma, \Sigma^\prime)
&=\langle \mathcal{T} T^{\mu a}(\Sigma) , T^{\rho b}(\Sigma^\prime) \rangle
= \frac{4}{\sqrt{-e(x)}\,\sqrt{-e(y)}}
\frac{\delta^2 F[e,\omega]}{\delta e^a_{\mu}(x) ,\delta e^b_{\rho}(y)} \\
\mathcal{L}^{\mu a,\rho b c}(\Sigma, \Sigma^\prime)  
&=\langle \mathcal{T}  T^{\mu a}(\Sigma) , \hat{S}^{\rho b c} (\Sigma^\prime) \rangle
= \frac{4}{\sqrt{-e(x)}\,\sqrt{-e(y)}}
\frac{\delta^2 F[e,\omega]}{\delta e^a_{\mu}(x)\,\delta \omega^{bc}_{\rho}(y)} \\
\mathcal{S}^{\mu a b , \rho c d}(\Sigma, \Sigma^\prime)
&= \langle \mathcal{T}  \hat{S}^{\mu a b} (\Sigma), \hat{S}^{\rho c d} (\Sigma^\prime) \rangle = \frac{4}{\sqrt{-e(x)}\,\sqrt{-e(y)}}
\frac{\delta^2 F[e,\omega]}{\delta \omega^{a b}_{\mu} (x)\,\delta \omega^{c d}_{\rho}(y)} .
\end{align}
where $\mathcal{T}$ is the time ordering operator. The consequence is that we can construct the Green's functions that satisfy \eqref{WI}. 

After some algebra, detailed in the Appendix, we get the Ward identities, in the $K^{ab}_\mu \rightarrow 0$ limit
\begin{align}  
\partial_\mu  G^{\mu a,\rho b} (\Sigma,\Sigma^\prime)  =&  - \frac{1}{\sqrt{-e}}\delta (\Sigma - \Sigma^\prime)\Bigg[  \partial_\mu \Big( e^{\mu a} \langle T^{\rho b } \rangle +  e^{\rho a} \langle T^{\mu b} \rangle - \eta^{a b} \langle T^{\mu \rho } \rangle
\Big) -  \Big( \eta^{\mu \alpha } \langle T^{ \rho \beta } \rangle  + \label{2point1}\nonumber \\
& \eta^{\mu \beta } \langle T^{\rho \alpha } \rangle  - \eta^{\alpha \beta } \langle T^{\mu \rho} \rangle \Big) ( e^a_{\alpha}\delta_{\mu\beta}
+ e^a_{\beta}\delta_{\mu \alpha} ) \Bigg] \\ 
\partial_\mu \mathcal{L}^{\mu a,\rho b c}(\Sigma, \Sigma^\prime) =& - \frac{1}{\sqrt{-e}} \delta(\Sigma - \Sigma^\prime) \bigg[ \langle T^{\rho a} \rangle \eta^{b c}  + \partial_\mu \Big( \omega_{\mu a c} \langle T^{\rho b } \rangle +  \omega^{\rho a b } \langle T^{\mu c} \rangle - \eta^{a b} \langle T^{\mu \rho } \rangle \Big) \bigg]  \\
\partial_\mu \mathcal{S}^{\mu a b,\rho c d } (\Sigma,\Sigma^\prime) 
=& - \frac{1}{\sqrt{-e}}\delta (\Sigma - \Sigma^\prime) \Bigg[  \partial_\mu \Big( \omega^{\mu a b}\langle S^{\rho c d} \rangle -  \omega^{\rho a d}  \langle S^{\mu b c} \rangle   \Big) -  \partial_\mu  \Big(  \langle S^{\mu \alpha \beta} \rangle +  \langle S^{\alpha \mu\beta } \rangle  -   \nonumber \\ 
&  \langle S^{\beta \mu\alpha} \rangle \Big) ( e^c_{\alpha}\delta_{\beta}^{\rho}
+ e^c_{\beta}\delta_{\alpha}^{\rho} ) \eta^{a b}   +  \omega_\mu^{a d} \eta^{bc}  \Big( \langle T^{\mu \rho} \rangle - \langle T^{\rho \mu} \rangle \Big) \Bigg] 
\label{2point3}
\end{align}


Such a set of equations as \wardid can provide a way to evolve a Gaussian partition function in time following the reasoning in \cite{torr1}. In this sense, $\mathcal{Z}$ which is being differentiated in \eqn{parttorsion} has the form \cite{becspin}.  According to the approach in this work, pseudogauge transformations should redistribute angular momentum between $\Omega_{\mu \nu \alpha}$ and $S_{\mu \nu \alpha}$ leaving $\lnz$ unchanged.  This can be consistently realized within the Gaussian ansatz.

Before continuing a caveat is in order:  To ensure a causal evolution from the Ward identity, the weak energy condition ($WEC$) \cite{Visser} and conjecture $\forall d\Sigma_\mu,T_{\mu \nu}d\Sigma^\mu d\Sigma^\nu \geq 0$ since every observer must measure a non-negative energy density. However, the effect of spin  modifies the structure of stress tensor, and consequently, the restrictions holding for $WEC$ appears non-trivial and dependent of spin frame-choice. 
One has to remember,however, that the dynamics is pseudo-gauge independent.  One should therefore be able choose a pseudo-gauge where all angular momentum is in the vorticity (this essentially involves choosing a foliation where spin is isotropic).  In this pseudogauge only $T_{\mu \nu}$ is non-trivial and we should recover the usual WEC. Thus $T^{\mu\nu}_{BEL} u_\mu u_\nu \geq 0$, where $T^{\mu\nu}_{BEL}$ is the Belinfante–Rosenfeld tensor.  This is essentially what is done in \cite{pseudogauge2}.

This pseudo-gauge, however, would most likely not be the best for quantitative calculations: Wile $Q_{...},W_{...}=0$, $C_{\alpha \beta \mu \nu}(\Sigma,\Sigma')\rightarrow  C_{\alpha \beta \mu \nu}(J^{\mu\nu}_{\Sigma,\Sigma'})$ be dependent on angular momentum about the axis defined by $\Sigma,\Sigma`$ , a highly non-local quantity (see \eqn{eq:angmomcons}).   Thus, $d\Sigma$ and $d\Sigma'$ in \eqn{partfull} would strongly correlate, and the resulting dynamics, while causal would have strong long-distance correlations (in \cite{torr1} $\Sigma$ and $\Sigma'$ should factorize as all conservation laws are local).   In such an approach a gradient expansion would be inappropriate, as indeed is evident from \cite{pseudogauge2} where no spin gradient terms appear in the constitutive relations.

A pseudo-gauge where $\ave{S}_{\mu\nu \alpha \beta},Q_{\mu \nu \alpha},W_{\mu \nu \alpha \beta}\ne 0$ has more parameters, but these parameters can be thought of as local fields of lagrange multipliers to enforce these non-local conservation laws.  This approach, paralleling that of spin chemical potentials \cite{spinpot1,spinpot2,spinpot3} could lead to an easier to handle partition function in practice, and the dynamics in this limit is most likely similar, close to the low viscosity limit, to that defined in \cite{spin2,spin3}.   This is analogous to the choice between the noise in Israel-Stewrt type and BDNK type hydrodynamics \cite{hipnoise}:  The former has more degrees of freedom but more localized noise, the latter has non-local correlations due to the choise of flow being subject to non-local constraints. If the dissipative terms are included in the correlator, these can be thought of as different foliations, analogous to the setting of a gauge \cite{torr1}

  By inspection of \eqn{partfull} together with \wardid it is clear that
\begin{equation}
C^{\mu \nu \alpha \rho}=\left( G^{\mu a,\rho b} e_a^\alpha  e_b^\beta \right)^{-1} \eqcomma Q^{\mu \rho \alpha \beta \gamma}=\left( \mathcal{L}^{\mu a,\rho b c} e_a^\alpha e_b^\beta e_c^\gamma\right)^{-1} \eqcomma W^{\mu \nu \alpha \beta \gamma \rho} = \left( \mathcal{S}^{\mu a b,\rho c d } e_a^\nu e_b^\alpha  e_c^\beta e_d^\gamma\right)^{-1}
\end{equation}
Where $\left( A_{\mu \nu_1 ... \alpha \alpha_1}\right)^{-1} A^{\mu \mu_2... \alpha \alpha_2}=\delta_{\mu_1}^{\mu_2}\times ... \times \delta_{\alpha_1}^{\alpha_2}$.

The entropy content in a fluctuation scale ''cell'', in analogy with \cite{torr1} is
\begin{equation}
\label{crooks}
\Delta \lnz= C_{\mu \nu \alpha \rho}\Delta^\mu \beta^\nu \Delta^\alpha \beta^\rho+Q_{\mu \nu \zeta \gamma \rho}\Delta^\zeta (\beta_\delta S^{\rho \gamma \delta})\Delta^\mu \beta^\nu + W_{\mu \nu \rho \gamma \omega \zeta} \Delta^\zeta (\beta_\xi S^{\mu \nu \xi})\Delta^\omega (\beta_\iota S^{\rho \gamma \iota})
\end{equation}
A Crooks theorem based algorithm \cite{crooks,hipcrooks} where the probability configurations jump according to their entropy content
\begin{equation}
\frac{P\left(\mathcal{A}+\Delta \mathcal{A}|\mathcal{A}\right)}{P\left(\mathcal{A}|\mathcal{A}+\Delta\mathcal{A}\right)}=\exp\left[\Delta \lnz\right] 
\end{equation}
(Here $\mathcal{A}$ is the combined 
\[\ \mathcal{A}=\ave{T_{\mu \nu}(\Sigma)},\ave{S_{\mu \nu \alpha}(\Sigma)},C_{\mu \nu \alpha \rho}(\Sigma),Q_{\mu \nu \zeta \gamma \rho}(\Sigma),W_{\mu \nu \rho \gamma \omega \zeta}(\Sigma)\]
an algorithm choosing $\Delta \mathcal{A}$ on the basis of the Ward identities \wardid and accepting/rejecting on the basic of \eqn{crooks} could evolve an approximately Gaussian ensemble of configurations in a generally covariant way from an initial condition such as
\cite{initial}.  Note that for small systems \cite{cmslocal} such initial conditions are typically not pseudo-gauge invariant \cite{tmd}, but the dynamics described here should nevertheless describe an evolution where initial conditions defined in different pseudo-gauges should evolve in equivalent ways up to a pseudo-gauge reparametrization.    

This definition of local equilibrium also accomodates the evolution to global equilibrium in the presence of metastable states.  If a vortical structure is not a global minimum of the free energy, fluctuations will eventually break it apart as the system evolves.   Thus, negative moments of inertia states, found in some lattice simulations \cite{chernneg}, will generally not survive for long times.

Equivalently, we can write down this dynamics using linear response theory \cite{torr1,forster}, since they provide three more constraints in addition to  \wardid.  In analogy with \cite{torr1}
\textcolor{black}{
\begin{align}\label{Tuv} 
& \ave{\hat{T}^{\mu\nu}(\Sigma_o)} = \int d \Sigma_0^\prime \exp{ [ i\epsilon \Sigma^\prime_0]} \left[ G^{\mu\nu\alpha\beta} (\Sigma_0 - \Sigma^\prime_0) \delta g_{\alpha\beta} + \mathcal{L}^{\mu\nu \alpha \beta \gamma } (\Sigma_0 - \Sigma^\prime_0) \delta \Xi_{\alpha \beta \gamma} \right]  \\  \label{Suv}
&\ave{\hat{S}^{\mu \nu \gamma}(\Sigma_o)} = \int d \Sigma_0^\prime \exp{ [ i \epsilon \Sigma^\prime_0 ] } \left[ \mathcal{L}^{\mu \nu \gamma \alpha \beta } (\Sigma_0 - \Sigma^\prime_0) \delta g_{\alpha \beta} + \mathcal{S}^{\mu \nu \gamma \alpha \beta \rho} (\Sigma_0 - \Sigma^\prime_0)  \delta \Xi_{\alpha \beta \rho} \right],
\end{align}}
where the vorticity can be obtained by subtracting the spin from the angular momentum content, which is conserved.

By disturbing the sytem from the full thermodynamics equilibrium, it is useful to consider the system evolves adiabatically. In such approach, we evaluate the one-point correlation of $T^{\mu\nu}$ and $S^{\mu \nu \lambda}$ from the variation of the metric $g_{\mu\nu}$ and gauge field $\Xi_{\alpha\beta\gamma}=\omega^{bc}_\beta e_{b \alpha} e_{c\gamma} $ by turning on  the source adiabatically 
\begin{equation}
\left. \begin{array}{ll} \{ \delta g_{\mu \nu}, \delta \Xi_{\alpha \beta \gamma} \} \times e^{\epsilon \Sigma_0} & \Sigma_0>0\\
0 & \Sigma_0 \leq 0
\end{array}
\right.
\end{equation}
where $\epsilon$ is an infinitesimal positive constant.
This allows us to extend the procedure used in section IV of \cite{torr1} (see also \cite{forster})  to get
\textcolor{black}{
\begin{align}\label{Tuvcorr} 
& \ave{\hat{T}^{\mu\nu}(\Sigma_o)}_{\Sigma + d \Sigma } = \int d \Sigma_0^\prime  \left[ \breve{G}^{\mu\nu\alpha\beta} (\Sigma_0 - \Sigma^\prime_0) \ave{\hat{T}_{\alpha\beta}(\Sigma_o')} + \breve{\mathcal{L}}^{\mu\nu \alpha \beta \gamma } (\Sigma_0 - \Sigma^\prime_0) \ave{\hat{S}_{\alpha \beta \gamma}(\Sigma_o')} \right]  \\  \label{Suvcorr}
&\ave{\hat{S}^{\mu \nu \gamma}(\Sigma_o)}_{\Sigma + d \Sigma }  = \int d \Sigma_0^\prime  \left[ \breve{\mathcal{L}}^{\mu \nu \gamma \alpha \beta } (\Sigma_0 - \Sigma^\prime_0)\ave{\hat{T}_{\alpha \beta}(\Sigma_o')} + \breve{\mathcal{S}}^{\mu \nu \gamma \alpha \beta \rho} (\Sigma_0 - \Sigma^\prime_0)  \ave{\hat{S}_{\alpha \beta \rho}(\Sigma_o)'} \right],
\end{align}}
where $\breve{f}(\Sigma)$ is defined in terms of the Fourier transform in the space-like directions $\tilde{f(\Sigma_0,\vec{k})}$ \cite{forster,torr1} as
\begin{equation}
\breve{f}(\Sigma^\mu) = \frac{1}{2i \pi} \lim_{\epsilon \rightarrow 0} \left[\int \frac{1}{\sqrt{-g}}d^{3} k e^{-ik_0 x^0} \left(\frac{\tilde{f}\left(\Sigma_0,\vec{k}\right)}{\tilde{f}\left(-i\epsilon \Sigma_0,\vec{k}\right)}-1\right)\right]
\end{equation}
Looking at \wardid it is clear that \eqn{Tuvcorr} and \eqn{Suvcorr} are pseudo-gauge covariant.  While each component of the equations change with the pseudo-gauge, the Ward identities are unaffected, which means neither is the dynamics.   
However, \eqn{Tuvcorr} and \eqn{Suvcorr} are manifestly not covariant.
However, as shown in \cite{torr1} the covariance can be restored if we assume the configuration space degrees of freedom of the system have in addition volume preserving diffeomorphisms, the sysmmetries of ideal hydrodynamics \cite{torr2}.   Given that the dynamics is pseudo-gauge independent, choosing the Belinfante-Rosenfeld pseudo-gauge extends the proof of covariance in \cite{torr1} to the spin case.   Note also that linear respose is inherently causal.  If only the evolution of the average quantities is tracked, the extra degrees of freedom  of $\ave{S_{\mu \alpha \beta}}$ in the co-moving frame accomodate the non-equilibrium spin demanded in \cite{spin3}, keeping the theory causal.

The form of \eqn{partfull} also illustrates why spin is a ''slowly equilibrating variable \cite{tors3} although it is not associated with a conserved quantity and is in fact related to a fluctuation scale rather than a dissipation one \cite{rybreview}.   The point is that pseudogauge invariance requires a partition function where the angular momentum is separated from the rest of the energy-momentum tensor.  Fluctuation dissipation then requires two time-scales, which in principle are of the same order, thus ensuring a ''long'' spin relaxation time.  Causality constraints 
\cite{spin3,spin4} provide further lower limits to the relation between these quantities.
On the other hand, In analogy to \cite{torr1}, fluctuations that "just change the pseudogauge" are not physical fluctuations but are ghost-like reparametrizations. This has the potential to dramatically improve the applicability of hydrodynamics to the average evolution in strongly coupled systems with few degrees of freedom.

The physical picture of why dynamics in this approach is invariant under pseudogauge shifts parallels \cite{torr1}.    Pseudogauge redistributes conserved angular momentum between orbital angular momentum and spin.   So do changes in foliation, $d\Sigma_\mu$.   Thus, choosing $d\Sigma_\mu$ in the context of hydrodynamics with spin is equivalent to choosing a pseudo-gauge.  Observables will depend on such a choice but dynamics should not, in a similar way that flow velocity $u_\mu$ and 4-temperature $\beta_\mu$ are ambiguus in dissipative hydrodynamics.   The fact that pseudo-gauge symmetry is equivalent to a non-inertial coordinate transformation coupled to a field redefinition \cite{brauner} also makes it clear that for dynamics to be pseudo-gauge independent it must include at least two cumulants (i.e. it is not just an average theory), which indeed here we do.   Partition functions, either quantum or statistical, are generally invariant under field redefitions \cite{weinberg,equiv1}, but truncations based on evolution equations of operator averages generally break this invariance.  Just like in Dyson-Schwinger equations\cite{weinberg}, the redefinition invariance  appears as a constraint between n-point functions, which can be enforced in local equilibrium \cite{torr1}.

In conclusion, we have argued that keeping hydrodynamic fluctuations non-perturbative via a Gaussian ansatz and evolving them via a Ward identity can give a pseudo-gauge independent spin hydrodynamics, analogously to the frame-independent hydrodynamics of \cite{torr1}.   While numerical simulations with this approach are a long way away, it provides some useful conceptual insights, such as the role of the spin chemical potential as opposed to approaches focusing on total angular momentum, and clarifying the scale of spin equilibration.  For spins higher than 1/2, this approach should also be Gauge indepedent, although in this case the dynamics is likely to be very different from naive expectations, due to the presence of Gauge redundancies \cite{ghosts}.   We hope that a future numerical implementation, combined with continuing investigation of spin hydrodynamics, particularly in small systems \cite{spin1,cmslocal}, will lead to quantitative tests of this approach.

\textbf{Acknowledgements} GT started thinking about these issues thanks to the fruitful discussions with, and hospitality of, Umut Gursoy in february 2022.
Unfortunately, GT took years to properly understand the key ideas of \cite{tors1,tors2} on which this work is founded.   GT very much regrets that Umut Gursoy's tragic and untimely passing prevented him from participating in this work, which is dedicated to him.  GT
thanks Bolsa de produtividade CNPQ 305731/2023-8 and FAPESP 2023/06278-2 as well as
participation in the tematico 2023/13749-1 for support.
\appendix
\section{Supplemental material: Details of the derivation of the Ward identity}

We here discuss the effective action, $ F = - \ln{\mathcal{Z}[\mathcal{J}^\mu,\mathcal{A}^\mu]}$ with the pair relations $\mathcal{J} \equiv \{ T_{\mu}^{a},S_{\mu}^{ab} \}$ and $\mathcal{A}\equiv \{ e_{\mu}^{a},\omega_\nu^{ab} \}$. A continuous symmetry of the action leads to constraints on correlation functions. Using the properties of time-ordering and the conservation of the current, we have
\begin{equation}\label{WI}
\partial_\mu \langle \mathcal{T}  \, \mathcal{J}(x) \, \mathcal{A}(x_1) \dots \mathcal{A}(x_n) \rangle_{(e,\omega)}
= - \sum_{i=1}^{n} \delta^{(4)}(x - x_i)
\langle \mathcal{T} \, \mathcal{A}(x_1) \dots \mathcal{A}(x_n) \rangle.
\end{equation}
which relate different n-point correlation functions in the general background metric $\{ e_\mu^a, \omega_\mu^{ab}\}$. We note that in deriving the eq. \eqref{WI}, the contact terms happen when the spacetime coordinates coincide (the two insertions approach the same point), so the product of operators becomes singular and the correlation function becomes a distribution through the Ward Identities (symmetry $\rightarrow$ relations between correlation functions). In general terms, since the free energy $F = - \ln Z$ turns out to be important in the analysis of the thermodynamic properties, we can most conveniently construct the perturbation expansion in full analogy with the quantum field theory. The correlation functions are developed in a Taylor series because we are limiting to small deviations
\begin{align}
& F[e_\mu^a,\omega^{ab}_\mu,h] = F_o [e_\mu^a,\omega^{ab}_\mu] \ + \nonumber \\ 
 & \sum_{n=1}^{\infty} \frac{1}{2^n\,n!} \int d^4 x_1 \dots d^4 x_n \, \sqrt{-e} \dots \sqrt{-e}  \mathcal{H}_{a_1\dots a_n}^{\mu_1 \dots\mu_n}(x_1,\dots,x_n; \mathcal{A})\,
h^{\mu_1}_{a_1 b_1} (x_1) \dots h^{\mu_n}_{a_n b_n} (x_n)   
\end{align}
where $F_o$ is unperturbed. Each term of the Taylor expansion series is defined by 
\begin{equation}
\mathcal{H}_{a_1\dots a_n}^{\mu_1 \dots\mu_n}(x_1,\ldots,x_n; A_\mu) \equiv \frac{2^n}{\sqrt{-e(x_1)}\dots\sqrt{-e(x_n)}}\ 
\frac{ \delta^{n} F [e,\omega]}{\delta \mathcal{A}_{1} (x_1) \dots \delta \mathcal{A}_{n} (x_n)}\label{Tn}
\end{equation}
where $\mathcal{H} \equiv \{ G,\mathcal{L}, \mathcal{S} \} $. The infinitesimal transformtaion of the vielbein and spin connection following the diffeomorphism and pseudogauge invariance
\begin{equation}\begin{cases}
\delta_\chi e^a_\mu  = \xi^\nu \partial_\nu e^a_\mu  + e^a_\nu  \partial_\mu \xi^\nu - \alpha^a_b  e^b_\mu  ,\\  
\delta_\chi \omega^a_{\mu b}  = \xi^\nu \partial_\nu \omega^a_{\mu b }  + \omega^a_{\nu b }  \partial_\mu \xi^\nu 
+ \partial_\mu \alpha^a_b - \alpha^a_c \omega^c_{\mu b}  
+ \alpha^c_b \omega^a_{\mu c} ,
\end{cases}\end{equation}
where $\chi \equiv \{ \xi^\mu, \alpha^a_b \} $ is a set of local infinitesimal parameters responsible for diffeomorphism ($\xi^\mu$) local Lorentz transformation ($\alpha_{ab} = - \alpha_{ba}$). Introducing the indentity $g_{\mu\nu} = e^a_\mu e^b_\nu \eta_{a b}$, the variation of the metric induced by vierbein is 
\begin{equation}
\delta g_{\mu\nu} = \eta_{ a  b } 
\big( e_\nu^{  b } \delta e_\mu^{  a } + e_\mu^{a} \delta e_\nu^{b}\big)  \qquad \frac{\delta g_{\mu\nu}(x)}{\delta e^a_{\ \rho}(y)} = ( e_{a\nu}\delta^\rho_\mu + e_{a\mu}\delta^\rho_\nu )\delta(x-y)
\end{equation}
where $\frac{\partial e^a_{\ \mu}}{\partial e^c_{\ \rho}}
= \delta^a_c \delta^\rho_\mu$. The linear variation of the Christoffel connection is
\begin{equation}
\delta \Gamma^\rho_{ \mu\nu} = 
e_{a}^{ \rho} \partial_\mu \delta e_\nu^{ a} - \Gamma^\lambda_{ \mu\nu} e_{a}^{ \rho} \delta e_\lambda^{ a} 
+ \omega_{\mu b}^{a} e_{a}^{ \rho} \delta e_\nu^{ b} + e_\nu^{ b} e_{a}^{ \rho} \delta \omega_{\mu b}^{ a}.
\end{equation}

We should note that just like the averages are not pseudogauge independent, neither are correlators.
When we plug the pseudo-gauge transofmraitons into the LHS of \wardid we get
\begin{eqnarray}
\partial_\mu  G^{\mu a,\rho b} (\Sigma,\Sigma^\prime) &\rightarrow & \partial_\mu  G^{\mu a,\rho b} (\Sigma,\Sigma^\prime) + \partial_\xi ( \delta(\Sigma-\Sigma^\prime)   (\phi^{\xi, \beta \rho} +\phi^{\beta,\xi\rho}+\phi^{\rho,\xi \beta}) )  \\
\partial_\mu \mathcal{L}^{\mu a,\rho b c}(\Sigma, \Sigma^\prime) &\rightarrow& \partial_\mu \mathcal{L}^{\mu a,\rho b c}(\Sigma, \Sigma^\prime) - \partial_\xi ( \delta(\Sigma-\Sigma^\prime)  ( \phi^{\xi, \beta \rho} + \phi^{\rho,\xi \beta} ) ) e_\beta^c \eta^{ab} \\ 
\partial_\mu \mathcal{S}^{\mu a b,\rho c d } (\Sigma,\Sigma^\prime) &\rightarrow& \partial_\mu \mathcal{S}^{\mu a b,\rho c d } (\Sigma,\Sigma^\prime) +  \partial_\mu ( \delta(\Sigma-\Sigma^\prime) \phi^{\rho, \alpha \beta} e^{\mu a} ) e^c_{\beta} \omega_\alpha^{bd}   - \partial_\mu ( \delta(\Sigma-\Sigma^\prime)   \phi^{\mu, \alpha \beta} e^{\rho a} ) \times  \nonumber\\
&&  e^b_{\beta} \omega_\alpha^{c d} + \partial_\mu ( \delta(\Sigma - \Sigma^\prime) (\phi^{\mu, \alpha \rho} + \phi^{\alpha, \mu \rho} + \phi^{\rho, \mu \alpha} )) e^c_{\alpha} 
+ 2 \partial_\xi ( \delta(\Sigma - \Sigma^\prime) \phi^{\xi, \mu \rho } )\omega_{\mu}^{a d}\eta^{b c} \nonumber \\
 \end{eqnarray}
the pseudo-gauge symmetry is in the relations between correlators and contact terms,not in each term separately.

Thus the thermodynamics is specified by an internal angular and spin momentum. 

\begin{equation}
   \langle P_s ^2 \rangle - \langle P_s \rangle^2 = \chi_s  k_b \Omega^2  \Rightarrow  \langle \Delta S^{\mu \nu \alpha} \Delta S_{\rho \lambda \beta} \rangle \sim \left. \frac{\partial^2 \ln \mathcal{Z}}{\partial \Omega_{ \nu \alpha} \partial \Omega_{\lambda \beta} } \right|_0 d \Sigma_\mu d \Sigma_\rho 
\end{equation}
where $P_s$ and  $\chi_s$ are the roational viscosity and density of spin momentum, respectively. We can write down the new ``Helmholtz`` free energy $ - d^2 \ln{\mathcal{Z}}/ (dT d \Omega)$ which gives to the Maxwell relation $ dS / d \Omega = d L / dT$, with $L$ the angular momentum. From the partition function, we 
\begin{equation}
   \langle E^2 \rangle - \langle E \rangle^2 = D_s  k_b T^2  \Rightarrow  \langle \Delta S^{\mu \nu \alpha} \Delta T_{\rho \lambda} \rangle \sim \left. \frac{\partial^2 \ln \mathcal{Z}}{\partial \Omega_{ \nu \alpha} \partial \beta_{\lambda} } \right|_0 d \Sigma_\mu d \Sigma_\rho 
\end{equation}
Energy fluctuations $\rightarrow$ spin fluctuations $\rightarrow$ transported by vortices. As the inclusion of spin introduces an anisotropy in the local equilibrium state.

The analytical continuation of the tensors $\mathcal{H}$ provides a way of generating the phenomenological transport coefficient that are, in general, anisotropic. Assuming the thermodynamic forces are considerably smaller than the characteristic length of macroscopic inhomogeneity.


\begin{equation}
D_s \sim \displaystyle\frac{\partial}{\partial \omega }\lim_{k \to 0} Im W_{t,x,t,x,t,x}  \eqcomma \chi_s \sim \displaystyle\frac{\partial}{\partial \omega }\lim_{k \to 0} Im Q_{x,y,x,-y}      
\end{equation}
where $\eta_r$ and $D_s$ are the transport coefficients of rotational viscosity (``resistence to rotation``) and spin diffusion (``stress due to spin gradient``), respectively. 
In the thermally asymptotic background these coefficients are analogous to the ones examined in \cite{spin4} and previous works.
It will also be interesting to see if this approach has a consistent zero-temperature quantum limit, to see if the TMD pseudo-gauge dependence seen, for example, in \cite{tmd} can be accounted a Gaussian wavefunction.


\begin{thebibliography}{10}


\bibitem{torr1}
G.~M.~Sampaio, G.~Rabelo-Soares and G.~Torrieri,
[arXiv:2504.17152 [hep-th]].

  \bibitem{spin1}
F.~Becattini and M.~A.~Lisa,
Ann. Rev. Nucl. Part. Sci. \textbf{70}, 395-423 (2020)
doi:10.1146/annurev-nucl-021920-095245
[arXiv:2003.03640 [nucl-ex]].

\bibitem{push}
J.~H.~Gao, G.~L.~Ma, S.~Pu and Q.~Wang,
Nucl. Sci. Tech. \textbf{31} (2020) no.9, 90
doi:10.1007/s41365-020-00801-x
[arXiv:2005.10432 [hep-ph]].

\bibitem{dirk}
N.~Weickgenannt, D.~Wagner, E.~Speranza and D.~H.~Rischke,
Phys. Rev. D \textbf{106} (2022) no.9, L091901
doi:10.1103/PhysRevD.106.L091901
[arXiv:2208.01955 [nucl-th]].

\bibitem{rybreview}
D.~Montenegro, R.~Ryblewski and G.~Torrieri,
Acta Phys. Polon. B \textbf{50} (2019), 1275
doi:10.5506/APhysPolB.50.1275
[arXiv:1903.08729 [hep-th]].

\bibitem{spin2}
D.~Montenegro, L.~Tinti and G.~Torrieri,
Phys. Rev. D \textbf{96}, no.5, 056012 (2017)
doi:10.1103/PhysRevD.96.056012
[arXiv:1701.08263 [hep-th]].

\bibitem{spinnonrel} 
R. F. Snider and K. S. Lewchuk
Journal of Chem. Phys. \textbf{46}, 3163 (1967); 10.1063/1.1841187


\bibitem{tors1}
A.~D.~Gallegos, U.~Gursoy and A.~Yarom,
JHEP \textbf{05}, 139 (2023)
doi:10.1007/JHEP05(2023)139
[arXiv:2203.05044 [hep-th]].

\bibitem{tors2}
A.~D.~Gallegos, U.~G{\"u}rsoy and A.~Yarom,
SciPost Phys. \textbf{11}, 041 (2021)
doi:10.21468/SciPostPhys.11.2.041
[arXiv:2101.04759 [hep-th]].

\bibitem{tors3}
M.~Hongo, X.~G.~Huang, M.~Kaminski, M.~Stephanov and H.~U.~Yee,
JHEP \textbf{11} (2021), 150
doi:10.1007/JHEP11(2021)150
[arXiv:2107.14231 [hep-th]].



\bibitem{spin3}
D.~Montenegro and G.~Torrieri,
Phys. Rev. D \textbf{100}, no.5, 056011 (2019)
doi:10.1103/PhysRevD.100.056011
[arXiv:1807.02796 [hep-th]].

\bibitem{spin4}
D.~Montenegro and G.~Torrieri,
[arXiv:2506.17327 [nucl-th]].


\bibitem{speranza2}
F.~Becattini, W.~Florkowski and E.~Speranza,
Phys. Lett. B \textbf{789} (2019), 419-425
doi:10.1016/j.physletb.2018.12.016
[arXiv:1807.10994 [hep-th]].

\bibitem{becspin}
F.~Becattini and R.~Singh,
[arXiv:2506.20681 [nucl-th]].


\bibitem{spinpot1}
W.~Florkowski, A.~Kumar and R.~Ryblewski,
Acta Phys. Polon. B \textbf{51} (2020), 945-959
doi:10.5506/APhysPolB.51.945
[arXiv:1907.09835 [nucl-th]].

\bibitem{spinpot2}
V.~V.~Braguta, M.~N.~Chernodub and A.~A.~Roenko,
Phys. Rev. D \textbf{111} (2025) no.11, 114508
doi:10.1103/xptn-qgfl
[arXiv:2503.18636 [hep-lat]].

\bibitem{spinpot3}
R.~L.~S.~Farias and W.~R.~Tavares,
Phys. Rev. D \textbf{112} (2025) no.5, L051902
doi:10.1103/5znc-7ztg
[arXiv:2507.08130 [hep-ph]].

\bibitem{zubbec}
F.~Becattini,
Lect. Notes Phys. \textbf{987} (2021), 15-52
doi:10.1007/978-3-030-71427-7{\_}2
[arXiv:2004.04050 [hep-th]].

\bibitem{pseudopush}
J.~H.~Gao, G.~L.~Ma, S.~Pu and Q.~Wang,
Nucl. Sci. Tech. \textbf{31} (2020) no.9, 90
doi:10.1007/s41365-020-00801-x
[arXiv:2005.10432 [hep-ph]].

\bibitem{pseudogaugebuz}
M.~Buzzegoli,
Phys. Rev. C \textbf{105}, no.4, 044907 (2022)
doi:10.1103/PhysRevC.105.044907
[arXiv:2109.12084 [nucl-th]].

\bibitem{flork}
A.~Das, W.~Florkowski, R.~Ryblewski and R.~Singh,
Phys. Rev. D \textbf{103}, no.9, L091502 (2021)
doi:10.1103/PhysRevD.103.L091502
[arXiv:2103.01013 [nucl-th]].

\bibitem{pseudogauge2}
F. Becattini and C. Hoyos,
[arXiv: 2507.09249 [nucl-th]].


\bibitem{jain}
J.~Armas and A.~Jain,
[arXiv:2601.14421 [hep-th]].



\bibitem{speranza}
N.~Weickgenannt, D.~Wagner and E.~Speranza,
Phys. Rev. D \textbf{105} (2022) no.11, 116026
doi:10.1103/PhysRevD.105.116026
[arXiv:2204.01797 [nucl-th]].

\bibitem{weinberg}
S.~Weinberg,
Cambridge University Press, 2005,
ISBN 978-0-521-67053-1, 978-0-511-25204-4
doi:10.1017/CBO9781139644167

\bibitem{misner}
C.~W.~Misner, K.~S.~Thorne and J.~A.~Wheeler,
W. H. Freeman, 1973,
ISBN 978-0-7167-0344-0, 978-0-691-17779-3

\bibitem{hehl}
F.~W.~Hehl,
Rept. Math. Phys. \textbf{9} (1976), 55-82
doi:10.1016/0034-4877(76)90016-1


\bibitem{noether}
S.~G.~Avery and B.~U.~W.~Schwab,
JHEP \textbf{02} (2016), 031
doi:10.1007/JHEP02(2016)031
[arXiv:1510.07038 [hep-th]].

\bibitem{jeonspin}
S.~Jeon,
EPJ Web Conf. \textbf{276} (2023), 01010
doi:10.1051/epjconf/202327601010
[arXiv:2310.11269 [nucl-th]].


\bibitem{brauner}
T.~Brauner,
Phys. Scripta \textbf{95} (2020) no.3, 035004
doi:10.1088/1402-4896/ab50a5
[arXiv:1910.12224 [hep-th]].

\bibitem{freese}
A.~Freese,
Phys. Rev. D \textbf{113} (2026) no.1, 016011
doi:10.1103/wmz3-lrbz
[arXiv:2506.04510 [hep-ph]].

\bibitem{ghosts}
G.~Torrieri,
Eur. Phys. J. A \textbf{56} (2020) no.4, 121
doi:10.1140/epja/s10050-020-00121-z
[arXiv:1810.12468 [hep-th]].

\bibitem{crooks}
G.~Torrieri,
JHEP \textbf{02} (2021), 175
doi:10.1007/JHEP02(2021)175
[arXiv:2007.09224 [hep-th]].

\bibitem{torr2}
G.~Torrieri,
Phys. Rev. D \textbf{109}, no.5, L051903 (2024)
doi:10.1103/PhysRevD.109.L051903
[arXiv:2307.07021 [hep-th]].



\bibitem{torr3}
T.~Dore, L.~Gavassino, D.~Montenegro, M.~Shokri and G.~Torrieri,
Annals Phys. \textbf{442}, 168902 (2022)
doi:10.1016/j.aop.2022.168902
[arXiv:2109.06389 [hep-th]].

\bibitem{diffusion}
G.~Torrieri,
[arXiv:2601.01656 [hep-th]].

\bibitem{grossi}
S.~Floerchinger and E.~Grossi,
Phys. Rev. D \textbf{105} (2022) no.8, 085015
doi:10.1103/PhysRevD.105.085015
[arXiv:2102.11098 [hep-th]].

\bibitem{kovner}
I.~I.~Kogan, A.~Kovner and J.~G.~Milhano,
JHEP \textbf{12} (2002), 017
doi:10.1088/1126-6708/2002/12/017
[arXiv:hep-ph/0208053 [hep-ph]].

\bibitem{gauss2}
Jena-Lasinio G.  Physics Reports.  \textbf{352(4-6)} 439-58 (2001)
[arXiv:cond-mat/0009219 [cond-mat]].


\bibitem{boulware}
S.~Deser and D.~Boulware,
J. Math. Phys. \textbf{8} (1967), 1468
doi:10.1063/1.1705368

\bibitem{heinz}
S.~Jeon and U.~Heinz,
Int. J. Mod. Phys. E \textbf{24}, no.10, 1530010 (2015)
doi:10.1142/S0218301315300106
[arXiv:1503.03931 [hep-ph]].


\bibitem{forster}
D. Forster, ”hydrodynamic fluctuations, broken symmetry and correlation functions”,
Addison-Wesley (1990)




\bibitem{probe}
Y.~Mao, M.~Tegmark, A.~H.~Guth and S.~Cabi,
Phys. Rev. D \textbf{76} (2007), 104029
doi:10.1103/PhysRevD.76.104029
[arXiv:gr-qc/0608121 [gr-qc]].

\bibitem{impediment}
A.~Bochniak, L.~D{\k{a}}browski, A.~Sitarz and P.~Zalecki,
Phys. Rev. Lett. \textbf{134} (2025) no.23, 231501
doi:10.1103/drdl-l2mp
[arXiv:2412.19626 [gr-qc]].

\bibitem{fabbri}
L.~Fabbri,
Universe \textbf{7} (2021) no.8, 305
doi:10.3390/universe7080305
[arXiv:1703.02287 [gr-qc]].


\bibitem{koivisto}
D.~A.~Gomes, J.~Beltr{\'a}n Jim{\'e}nez and T.~S.~Koivisto,
Phys. Rev. D \textbf{107} (2023) no.2, 024044
doi:10.1103/PhysRevD.107.024044
[arXiv:2205.09716 [gr-qc]].

\bibitem{carroll}
S.~M.~Carroll,
[arXiv:gr-qc/9712019 [gr-qc]].

\bibitem{pereira}
R.~Aldrovandi and J.~G.~Pereira,
Springer, 2013,
ISBN 978-94-007-5142-2, 978-94-007-5143-9
doi:10.1007/978-94-007-5143-9

\bibitem{shapiro}
I.~L.~Shapiro,
Phys. Rept. \textbf{357} (2002), 113
doi:10.1016/S0370-1573(01)00030-8
[arXiv:hep-th/0103093 [hep-th]].


\bibitem{Visser}
M. Visser and C. Barcelo, doi:10.1142/9789812792129 0014 [arXiv:gr-qc/0001099 [gr-qc]].

\bibitem{hipnoise}
L.~Gavassino, N.~Mullins and M.~Hippert,
Phys. Rev. D \textbf{109} (2024) no.12, 125002
doi:10.1103/PhysRevD.109.125002
[arXiv:2402.06776 [nucl-th]].

\bibitem{hipcrooks}
N.~Mullins, M.~Hippert and J.~Noronha,
Phys. Rev. Lett. \textbf{134} (2025) no.23, 232302
doi:10.1103/PhysRevLett.134.232302
[arXiv:2501.04637 [nucl-th]].

\bibitem{initial}
G.~Rabelo-Soares, G.~Vujanovic and G.~Torrieri,
[arXiv:2511.03851 [hep-ph]].


\bibitem{cmslocal}
A.~Hayrapetyan \textit{et al.} [CMS],
Phys. Rev. Lett. \textbf{135} (2025) no.13, 132301
doi:10.1103/6ywq-gm61
[arXiv:2502.07898 [nucl-ex]].


\bibitem{tmd}
K.~Fukushima and T.~Uji,
[arXiv:2603.11704 [hep-ph]].

\bibitem{chernneg}
V.~V.~Braguta, M.~N.~Chernodub, A.~A.~Roenko and D.~A.~Sychev,
Phys. Lett. B \textbf{852} (2024), 138604
doi:10.1016/j.physletb.2024.138604
[arXiv:2303.03147 [hep-lat]].

\bibitem{equiv1}
V.~Periwal,
Phys. Rev. Lett. \textbf{80} (1998), 4366-4369
doi:10.1103/PhysRevLett.80.4366
[arXiv:hep-th/9709200 [hep-th]].


\end{thebibliography}
\end{document}